# ADTOF: A LARGE DATASET OF NON-SYNTHETIC MUSIC FOR AUTOMATIC DRUM TRANSCRIPTION

**Mickaël Zehren**
Umeå Universitet
mzehren@cs.umu.se

**Marco Alunno**
Universidad EAFIT Medellín
malunno@eafit.edu.co

**Paolo Bientinesi**
Umeå Universitet
pauldj@cs.umu.se

## ABSTRACT

The state-of-the-art methods for drum transcription in the presence of melodic instruments (DTM) are machine learning models trained in a supervised manner, which means that they rely on labeled datasets. The problem is that the available public datasets are limited either in size or in realism, and are thus suboptimal for training purposes. Indeed, the best results are currently obtained via a rather convoluted multi-step training process that involves both real and synthetic datasets. To address this issue, starting from the observation that the communities of rhythm games players provide a large amount of annotated data, we curated a new dataset of crowdsourced drum transcriptions. This dataset contains real-world music, is manually annotated, and is about two orders of magnitude larger than any other non-synthetic dataset, making it a prime candidate for training purposes. However, due to crowdsourcing, the initial annotations contain mistakes. We discuss how the quality of the dataset can be improved by automatically correcting different types of mistakes. When used to train a popular DTM model, the dataset yields a performance that matches that of the state-of-the-art for DTM, thus demonstrating the quality of the annotations.

## 1. INTRODUCTION

Automatic drum transcription (ADT) consists of creating a symbolic representation of the notes played by the drums in a music piece. Two targets that would benefit from such transcriptions are musicians, for example when learning a musical piece, and music information retrieval (MIR), that can leverage the location of the notes to draw a deeper knowledge of a music track (e.g., its structure). ADT is known to be difficult to achieve. In fact, as explained in a recent state-of-the-art review by Wu et al. [1], it is tackled in several manners and with different levels of complexity. The most basic aspect undertaken is the automatic classification of isolated drum sounds. Here, we are interested in solving the more general and complex task of drum transcription in the presence of melodic instruments (DTM). In DTM, the input consists of polyphonic music (drums and accompanying instruments); the output is a log with time stamp and instrument for each drum note. As Wu et al. argued, much progress has been made recently in ADT (and, therefore, in DTM) thanks to deep learning approaches. However, a high volume of annotated data is needed for neural networks to perform well, and such data is difficult to obtain, mainly because the annotation process is labor-intensive. This explains why the publicly available datasets are usually either small (e.g. [2–4]) or consist of augmented data (e.g. [5,6]) or synthesized audio (e.g. [7,8]), both of which are not direct representations but only estimations of real music. Therefore, current datasets seem to be suboptimal for DTM either because of quantity or data authenticity.

In this work, we explore a way to reach both the required quantity and realism of the data needed for DTM by using crowdsourced annotations of a high volume of real-world (not synthetically created) audio tracks. In fact, we realized that this amount of data can be found in rhythm games such as RockBand [1] or PhaseShift [2]. In these games, one of the goals is to correctly play the drum line of a song on a toy drum kit. Songs come with the game, but players can also add audio tracks and their own annotations of the drums parts. Because of this feature, a large online community of players and musicians emerged to extend the catalog of playable tracks and share custom game files, also known as "custom charts". These data have the advantage of being fundamentally similar to the content of current ADT datasets and contain the audio source along with the representation of the notes being played on the drums.

The outcome of our work is a new methodology to build a dataset from custom charts. Following our method, we build a dataset named Automatic Drums Transcription On Fire (ADTOF) [3] that is composed of a large amount of realistic data. As mentioned above, the large amount is achieved through crowdsourcing annotations from a much larger group of people than previously observed, to our knowledge. Realism is due, instead, to the use of real-world as opposed to augmented or synthesized music tracks. Yet, quantity and realism are useful only if the annotations contain as few mistakes as possible.

---

[1] https://www.rockband4.com/
[2] https://store.steampowered.com/app/865250/Phase_Shift/
[3] The name is a reference to "Frets On Fire", one of the earliest rhythm game.

| Dataset | Hours | Classes | Real music |
|---|---|---|---|
| ENST [2] | 1.02 | 20 | ✓ |
| MDB [3] | 0.35 | 21 | ✓ |
| RBMA [4] | 1.72 | 24 | ✓ |
| SDDS [7] | 467 | 14 | ✗ |
| TMIDT [8] | 259 | 18 | ✗ |
| ADTOF (ours) | 114 | 5 | ✓ |

**Table 1**. List of datasets for DTM.

In order to ensure a sufficient quality, we used a systematic way of curating the data from an online source by selecting the tracks that are less likely to contain wrong annotations. In fact, while manually assessing the annotations, we found many discrepancies between the locations of the annotations and the positions of the actual sound onsets. To overcome this issue we adapted the automatic alignment technique described in the work of Driedger et al. [9] to correct the time precision of the annotations. We also found many inconsistencies in the labels used to designate specific instruments of the drum kit, which we solved by reducing the set of instrument classes to be detected. Finally, in order to assess how useful the annotations were after being processed, we evaluated our new dataset as both a training and test data for the popular convolutional recurrent neural network (CRNN) illustrated in the work of Vogl et al. [8]. The result is that ADTOF allows for the direct training of a model that achieves comparable performance to the state-of-the-art model trained on multiple other datasets. It also provides complementary information and generalization capability. [4]

The rest of this article is organized as follows: Section 2 contains a survey of related works. The data with the annotation and curation process are presented in Section 3, and the methodology to automatically clean them is detailed in Section 4. In Section 5, experiments on training and testing are presented. The results are then discussed in Section 6. Conclusions are drawn and future works are described in Section 7.

## 2. RELATED WORK

Multiple datasets with different characteristics have been created to solve specific aspects of ADT. Since we deal with DTM, though, in this section we discuss only datasets containing polyphonic music (see Table 1).

To our knowledge, the oldest public dataset suitable for DTM is ENST [2] created in 2006. This dataset contains the recordings of three professional drummers playing along with a variety of musical accompaniments composed for drum kit practicing. More recently, in 2017, MDB drums [3] was created by adding drum transcriptions to 23 tracks from MedleyDB [10], and RBMA [4] was released, with annotations, on the freely available album

---

[4] The set of tools developed to curate and process the annotations as well as the pre-trained models and complementary material can be found at https://github.com/MZehren/ADTOF

"Various Assets - Not For Sale: Red Bull Music Academy New York 2013". These datasets are standard in DTM, but they are limited in several ways. First, because of the difficulties inherent in annotating music, these datasets are small, with a cumulative time just above three hours. Second, the number of occurrences of each instrument in a drum kit is generally unbalanced, with some instruments (e.g., crash cymbal, ride cymbal) appearing much less than others (e.g., bass drum, snare drum). Lastly, in these datasets, data diversity is largely reduced (e.g., ENST contains audio from a limited number of drum kits, RBMA is biased toward a few music genres). As a consequence, the majority of DTM research [4, 5, 11–13] narrows down to the identification of three main drum classes — kick drum, snare drum and hi-hat.

As an effort to increase the size of the manual annotations, data augmentation was employed by Vogl et al. [5] and, more recently, by Jacques and Röbel [6]. In these studies, data augmentation techniques (e.g., pitch-shifting, time-stretching the audio) usually increased the performances of the model trained on the augmented data. However, according to some of the authors in a later work [8, p. 4], this improvement is limited.

Another approach taken to contrast data paucity is the generation of synthetic datasets which consist of synthesized audio generated from a symbolic representation of music (i.e. MIDI files). This technique allows to create larger datasets because it removes the labor needed to annotate the audio tracks, since the ground truth is deduced from the generation process. Moreover, audio synthesis gives the flexibility to balance instrument distribution by artificially replacing more common drum classes with sparser ones.

Following this approach, Cartwright and Bello proposed in 2018 the Synthetic Drum Dataset (SDDS) [7] that is multiple orders of magnitude larger than the previous datasets. In their work, the audio has been rendered from a collection of MIDI drum loops using randomly selected drum samples, augmented with harmonic background noise and other data augmentation methods. The same year, Vogl et al. created another synthetic dataset, which we refer to as TMIDT [8], by using MIDI files available online to synthesize both drums and the accompaniment parts in such a way that drums classes would be distributed in a natural and balanced fashion. Both these works indicate that models trained on a large synthetic dataset alone do not outperform models trained on small real-world datasets, with still possible performance improvements for some underrepresented classes when using TMIDT. Furthermore, Vogl et al. [8, p. 5] raised the concern that the atypical nature of drum patterns that underwent a balancing process could harm the model and they showed that this technique is ineffective when making evaluation on real-world datasets. In conclusion, results improve only when real data is somehow involved: by training with both synthetic and real data [7], or by training first with synthetic data and then refining the outcome with real data [8].

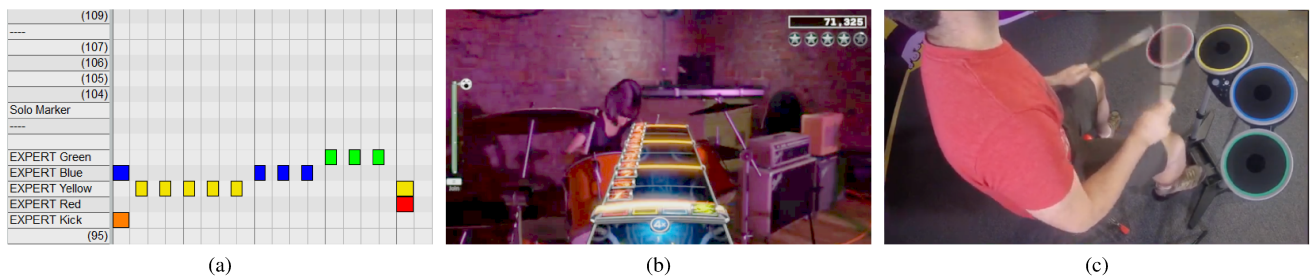

Figure 1. Digital audio workstation interface used to create the annotations with mouse and keyboard (a). The result is then playable in-game (b) using a drum-shaped game controller (c). Pictures from `http://docs.c3universe.com` and [14].

Even though the synthetic datasets are pushing further the state-of-the-art of DTM by providing more data, they do not seem to solve completely the need for more real annotated audio. In order to tackle this issue, we built the ADTOF dataset, the first to our knowledge that is both large and contains real-world audio tracks, with more than 114 hours of annotated music and the transcription of five different classes of drums.

## 3. DATASET

To build our dataset, we downloaded openly shared custom charts made for rhythm games. In this section, we discuss how the annotations were made and how we selected them.

### 3.1 Annotation process

The custom charts are created by players and musicians who desire to play along with their favorite tracks. They consist of the symbolic annotations of the drum onsets, usually in a MIDI file with standardized pitches representing the notes the player is supposed to play on the drums (we refer to them as GAMEPLAY annotations). In addition to the transcription, a chart contains beat information, the track's meta-data and the audio track to play along with. All this information is similar to what is found in an ADT dataset such as those introduced in Section 2. Some of the charts also contain other labels used to animate in-game musicians and are known as ANIMATIONS.

To create the annotations while ensuring quality and consistency in the data, guidelines and tools have been built by the players community and shared online.[5] These guidelines specify how to create the annotations in a digital audio workstation by manually building a grid of beats filled offline with mouse and keyboard (Fig. 1a). The result is then packaged with a tool and finally manually controlled in-game (Fig. 1b) and playtested on the game controller (Fig. 1c).

### 3.2 Data selection

Once the transcription is ready, it is shared on a collaborative website such as the popular "Rhythm Gaming World."[6] This specific website offers a listing of the transcriptions, a space for users' feedbacks and comments, and a forum to discuss and report mistakes. At the time of writing, Rhythm Gaming World lists more than 25,000 transcribed tracks. Most of them contain annotations for drums (but other instruments such as guitar, keyboards, and bass are also playable in rhythm games) and are labeled as "Pro drums", a tag added by the authors and meaning that the notes annotated should be all those found in the real song rather than an approximation or a simplification thereof. From this repository, we downloaded the top-rated 1700 tracks (the largest number of tracks we could fit into the local memory of the server used for the evaluation); this is the dataset we used in this study.

[5] `http://docs.c3universe.com/rbndocs/index.php`
[6] `https://rhythmgamingworld.com/`

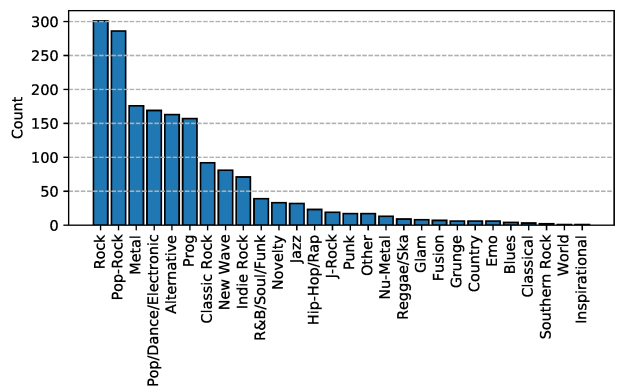

Figure 2. Genre distribution of tracks in ADTOF.

This subset contains an uneven distribution of music genres with a bias toward "Rock", possibly the most popular genre in the rhythm games community (see Fig. 2). The vast majority of the music is classified as Western with very few occurrences of "World", as the former is the most likely to contain parts for a typical drum kit like those used in rhythm games.

In order to use this subset of data, we created an open-source set of tools that automatically convert any custom charts into a standard format for ADT.

## 4. AUTOMATIC DATA CLEANSING

The annotations collected from custom charts are not directly usable for ADT. This is due in part to the crowd-sourced nature of the annotations, and in part to the specific target for which the annotations were created, that is, video games. On the one hand, different people will inevitably have different annotation expertise, and on the other hand, some of the annotations do not aim at being accurate, but at improving the players' experience. By conducting a visual and auditive inspection of random tracks in our dataset, we identified two recurring issues which could be detrimental for ADT. In this section we illustrate how such issues can be corrected automatically.

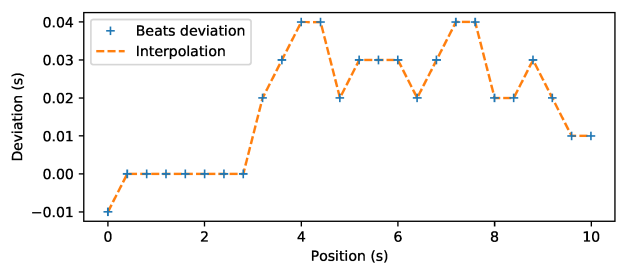

**Figure 3**. Deviation of the beats annotation as computed with the method from [9] and the interpolation used to correct the drum notes for the first 10 secs of a track.

### 4.1 Inaccurate timing

The first issue is a lack of time accuracy: We identified that many annotations are placed in a neighborhood of the actual notes. This might be due to human imprecision, but also to the fact that in some tracks the tempo is not constant (for instance because they come from live performances), making them especially difficult to annotate. The observed distance between real and annotated events might be too extreme for the training of a typical algorithm used for ADT. For example, the models discussed by Vogl [8] require the annotations to be roughly within 10 ms from the actual onset,[7] whereas we usually witnessed discrepancies around 50 ms.

To improve the alignment of the annotations, we adapt the work of Driedger et al. [9] whose idea is to correct human annotations of beats by "snapping" them onto their most likely position according to a beat tracking algorithm. The assumption behind this reasoning is that human annotations are meant to correspond to actual events but, due to human error, they are likely slightly misplaced; instead, beat tracking algorithms, if successful, locate beats accurately. Then, by comparing human annotations with algorithmically identified beats, it should be possible to correct wrongly placed human annotations. The beat tracking algorithm used is Böck's [15], available in the library Madmom [16]. We extend this initial method, originally meant to correct human taps recorded live, to also correct the annotated drum notes occurring in between the beats. To this end, we interpolate the corrections to intermediate positions as represented in Fig. 3. We use a linear interpolation as it is likely to represent the true deviation of the drum notes.

To further improve the overall quality of the dataset, we also include an automatic sanity check, to make sure that the majority of the corrected beats align to the algorithmically detected beats, and that the magnitude of the corrections does not exceed 80ms. A total of 140 tracks that do not satisfy these requirements were discarded.

---

[7] This is because those models are tuned to work at 100Hz and, thus, are making a prediction every 10 ms.

| Animation | Gameplay | ADTOF |
|---|---|---|
| Bass drum | Orange drum | BD |
| Snare drum | Red drum | SD |
| Rack tom 1 | Yellow drum | TT |
| Rack tom 2 | Blue drum | |
| Floor tom | Green drum | |
| Hi-hat open | Yellow cymbal | HH |
| Hi-hat close | | |
| Crash 1 | Blue and green cymbal | CY + RD |
| Crash 2 | | |
| Ride Cymbal | | |

**Table 2**. Classes used for the Animation and Gameplay annotations (presented in Sec. 3.1), and mapping onto ADTOF classes.

### 4.2 Inconsistent labeling

The second issue we identified in the annotations is the inconsistent use of labels. This issue is apparent for specific drums sounds such as the three variants of the toms (i.e., yellow, blue, and green drums), which are challenging to discriminate even within one track. Moreover, since the same variant of toms might sound drastically different depending on mix and playing style, it is especially difficult to achieve consistent labeling across tracks. Our solution consists in merging different classes into one (see Table 2, 2nd and 3rd columns), a simplification which is not uncommon in ADT, as the discrimination of the toms sounds is not as relevant as correctly identifying their presence.

Another cause of inconsistencies is the fact that some drums have an ambiguous representation in the toy drum kit used as game controller. For instance, we found that on the toy drum kit, the yellow in-game cymbal ambiguously represents both the open and closed hi-hat; similarly, the blue and green in-game cymbals are used interchangeably for crash and ride sounds. For a subset of tracks, these ambiguities can be resolved by looking at the Animation annotations (Table 2, 1st column), which are highly accurate and span a larger drumset. However, not all the tracks do have such annotations. In our dataset, we map the vocabularies used in Animation or Gameplay down to five classes, the largest number of classes for which most ambiguities are removed.

We also observed that in accordance with the guidelines for the annotators, to ease the gameplay, specific sounds that cannot be played on a toy drum kit, have been represented by combinations of drum hits (e.g., a snare flam is represented as a hit on both the red and yellow drum; an accentuated open hi-hat sound is played on the green cymbal). For these cases, we use the ANIMATION annotations, when available, to detect discrepancies with the GAMEPLAY annotations, and then we adopt the former as ground truth.

Finally, aware that the guidelines do not cover all possible cases and that not all annotators follow the guidelines, we performed one last check on the 10% tracks with the lowest prediction score according to a preliminary trained ADT algorithm. With this extra check, we removed a total of 88 tracks; among these, there were tracks containing multiple drums but with annotations for only one, and tracks with classes outside of the vocabulary wrongly annotated.

## 5. EXPERIMENTS

To evaluate the quality of our dataset (ADTOF), we use it in two typical tasks, namely the evaluation, and the training of a DTM model. In the first case, we compare the performance achieved by a state-of-the-art algorithm on ADTOF and on other datasets. Should the performance on ADTOF be worse than on the other datasets, then we would conclude that the ADTOF's annotations contain errors and/or that ADTOF's tracks are especially difficult to transcribe for the given algorithm. By constrast, a good performance on ADTOF indicates that the annotations and the algorithmic estimations are in good agreement with each other. Since it is very unlikely that annotations and estimations agree on mistakes, the agreement is a strong indication that both are correct. In the second case, one same model is trained (and tested) on different datasets. By evaluating the performance that the model trained on ADTOF achieves on other datasets, we can assess both the quality of ADTOF's annotations, and how representative ADTOF's tracks are for the task.

### 5.1 Model selection

The state-of-the-art model we evaluate and train on our dataset is the convolutional recurrent neural network (CRNN) presented in the work of Vogl et al. [4, 8]. This model achieves the best overall results according to the community evaluation MIREX[8] on their private data, is well studied, and it has been reproduced [7]. The strength of this model resides in the two types of building blocks used in its architecture, namely convolutional and recurrent layers (see Fig. 4). The convolutional layers model the local acoustic features of the onsets, while the recurrent layers model the temporal aspect of the drum patterns. This combination of types of layers has been found to give the best results.

The input of the neural network is a spectrogram depicting the variation of the frequencies over time. Specifically, due to its good performance with this model, *the log-frequency log-magnitude short-time Fourier transform* is used. A window size of 2048 samples and a hop size of

---
[8] Results for "RV3" at the URL https://www.music-ir.org/mirex/wiki/2018:Drum_Transcription_Results

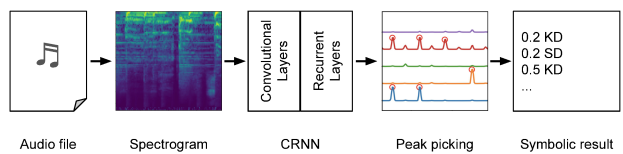

Figure 4. Overview of the automatic transcription.

441 samples are used for a target frame rate of 100Hz; the frequency bins are transformed to a logarithmic scale with 12 triangular filters per octave between 20 to 20,000Hz. The output of the network is an activation function displaying a value between zero and one for the five classes inferred. This value represents the confidence of the network that a note is played at this specific location. To extract a symbolic representation from these functions, a simple peak picking algorithm is used [17]. In the experiments, we carefully reproduced the whole procedure presented in Vogl's work and we implemented this network architecture with Tensorflow.

### 5.2 Evaluation

We compare results for ADTOF with three other datasets (ENST, MDB, RBMA) that contain real music (see Table 1) and by mapping their vocabulary onto our five classes. To carry out the comparison, we trained the CRNN on three different (combinations of) datasets, resulting in three different versions of the model. The first version is trained on ENST, MDB, and RBMA, and represents a baseline. The second version is first trained on TMIDT, and then on ENST, MDB, and RBMA; this is our reproduction of the state-of-the-art method. The last version is trained exclusively on ADTOF. In Fig. 5, these three versions are represented by the blue, orange, and green bars, respectively.

For each version, we followed a three-fold cross-validation strategy: testing is done on one split of each dataset while training and validation are done on the remaining splits of the dataset(s) used for training. This methodology ensures that the test data stays consistent across all models. When training on multiple datasets at the same time, we merged the corresponding training and validation sets together. In practice, for ENST, MDB, and RBMA, we used the three splits from [8] by iteratively selecting one as test data and further partitioning the remaining two as 15% validation and 85% training data. Whenever drum solo version of the tracks were available, they were added with their full mix version as additional training material, but not when testing. In contrast, since ADTOF is much larger, we partitioned the data in ten splits without overlap of artists between them to prevent leakage of information on similar-sounding tracks. The splits are used by iteratively selecting one as test data, one as validation and the remaining eight as training data.

The performance metric used is the well-known F-measure, computed with the package *mir_eval* [18] with a tolerance window of 50 $ms$ to stay consistent with previous research [1, 7]. We return values by counting the class specific true positives, false positives and false negatives across all tracks. We also compute an overall "SUM" value by counting across all tracks and all classes. This process is repeated three times on the different test sets of the cross-validations and we report the averaged results.

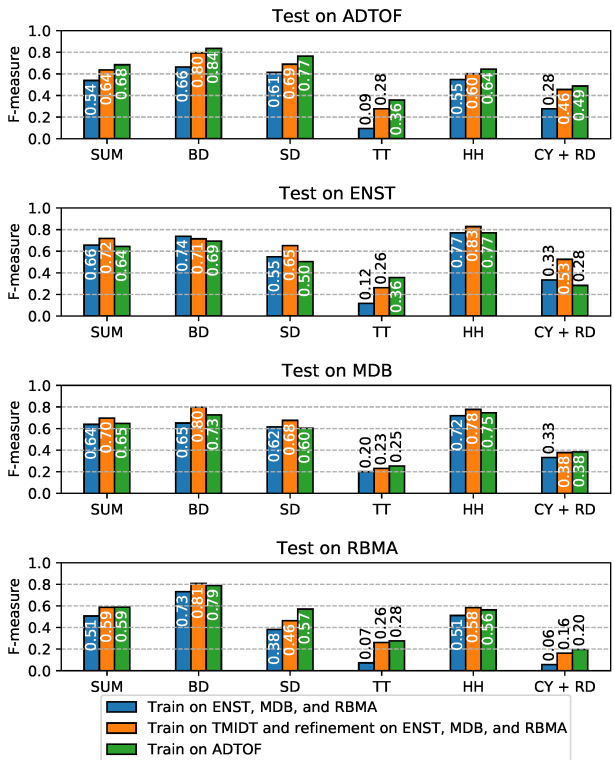

**Figure 5**. Plots representing the overall (SUM) and class-specific F-measure for multiple training and testing configurations.

## 6. RESULTS

The quality of the ADTOF dataset is assessed by (1) the performance that a state-of-the-art ADT algorithm achieves on it, and (2) the performance that a state-of-the-art ADT model trained on it achieves on multiple datasets. Fig. 5 depicts the results for both evaluations.

(1) In Fig. 5, the orange bars refer to the performance of the state-of-the-art algorithm, and different panels refer to different datasets. By comparing the orange bars in the top panel (ADTOF) and in the other three panels, we observe that the algorithm's performance on our dataset is entirely comparable with that on the other datasets. In detail, the classes toms (TT) and crash + ride cymbals (CY + RD) are difficult to estimate both in our and the other datasets. This fact is well understood, and typically linked to the sparsity of these classes in the training data. We note a slight increase in the snare drum (SD) performance on ADTOF compared to the other datasets, possibly meaning that the snare sounds in our tracks are more easily identified by the model. On the contrary, the hi-hat cymbal (HH) score is rather low, suggesting that the specific combinations of drum hits meant to ease the gameplay (see Sec. 4.2) are not fully corrected and harm the accuracy.

(2) We now focus on the performance achieved by one model when trained differently. The green bars refer to the model trained on ADTOF. Overall, our dataset trains the model equally well, and possibly better, than the current state of the art.

In detail, the model trained on ADTOF outperforms the baseline (blue bars) in the majority of the evaluations, indicating that the increased volume of the training data resulted in improved performance even on datasets unknown during training. Additionally, this fact suggests that the audio is representative of the task and enables generalization to other datasets. The model trained on ADTOF, and the model trained on TMIDT and refined on ENST, MDN, and RBMA, mostly outperform one another when evaluated on the same dataset(s) that was used for training; a few exceptions occur, in favor of the model trained on ADTOF. On the one hand, this is an indication that those datasets might be complementary, i.e., they all contribute useful information to the training. On the other hand, the exceptions in favor of ADTOF suggest that the inclusion of a larger number of real-world tracks results in better generalization capability of the model.

## 7. CONCLUSIONS

One of the main problems in ADT is to find a high volume of good quality data that the models can use during training. Unlike the most recent approaches that rely on synthesized and/or augmented audio to gather enough data, we proposed a new technique that leverages annotated tracks for rhythm video games. This source of data, to our knowledge unexplored until now in ADT, has the advantage of both containing authentic audio and being available online in a large quantity, thanks to the annotation process realized by a broad community of players. However, since the annotations originate from people with different expertise and are initially meant for a different purpose than ADT, they are not readily usable. Therefore, we illustrated a method to automatically select, transform, and clean the data. We further demonstrated that an algorithm trained on this dataset generalizes well and achieves state-of-the-art results, which is an indication that our method of gathering and curating the data is successful and should be further explored.

In future works, we are interested in evaluating repositories of custom charts other than "Rhythm Gaming World" in order to identify which of them is more suited to be used to build a dataset with our methodology. As an example, since the original tracks included in the video game RockBand are professionally annotated by the Harmonix gaming company, they may constitute a good quality set (because of better annotations) to explore.

## 8. REFERENCES

[1] C.-W. Wu, C. Dittmar, C. Southall, R. Vogl, G. Widmer, J. Hockman, M. Muller, and